\documentclass[twocolumn,twoside,slac_two]{revtex4}
\usepackage{graphicx}
\usepackage{fancyhdr}
\begin{document}

\title{ArgoNeuT: A Liquid Argon Time Projection Chamber Test in the NuMI Beamline}


\author{M. Soderberg, for the ArgoNeuT Collaboration}
\affiliation{Department of Physics, Yale University, New Haven, CT 06520, USA}

\begin{abstract}
Liquid Argon Time Projection Chamber detectors are ideally suited for studying neutrino interactions and probing the parameters that characterize neutrino oscillations.    The ability to drift ionization particles over long distances in purified argon and to trigger on abundant scintillation light allows for excellent particle identification and triggering capability.  In these proceedings the details of the ArgoNeuT test-beam project will be presented after a brief introduction to the detector technique.  ArgoNeuT is a 175 liter detector exposed to Fermilab's NuMI neutrino beamline.  The first neutrino interactions observed in ArgoNeuT will be presented, along with discussion of the various physics analyses to be performed on this data sample.
\end{abstract}

\maketitle

\thispagestyle{fancy}


%%%%%%%%%%%%%%%%%%%%%%%%%%%%%%%%%%
\section{Introduction}
Liquid Argon Time Projection Chambers (LArTPCs) are an appealing class of detectors that offers exceptional opportunities for studying neutrino interactions thanks to the bubble-chamber quality images they can provide.  The unique combination of position resolution, calorimetry, and scalability provided by LArTPCs make them a possible technology choice for future massive detectors.  There is an active program in the U.S. to develop LArTPCs, with the final goal of constructing a massive detector that can be used as a far detector in a long-baseline neutrino oscillation experiment.  The Argon Neutrino Teststand, or ArgoNeuT, project is an important early step in this program, and it will be the focus of this document.

%%%%%%%%%%%%%%%%%%%%%%%%%%%%%%%%%%
\section{LArTPC Technique}
\begin{figure}[h] %  figure placement: here, top, bottom, or page
   \centering
   \includegraphics[width=2in]{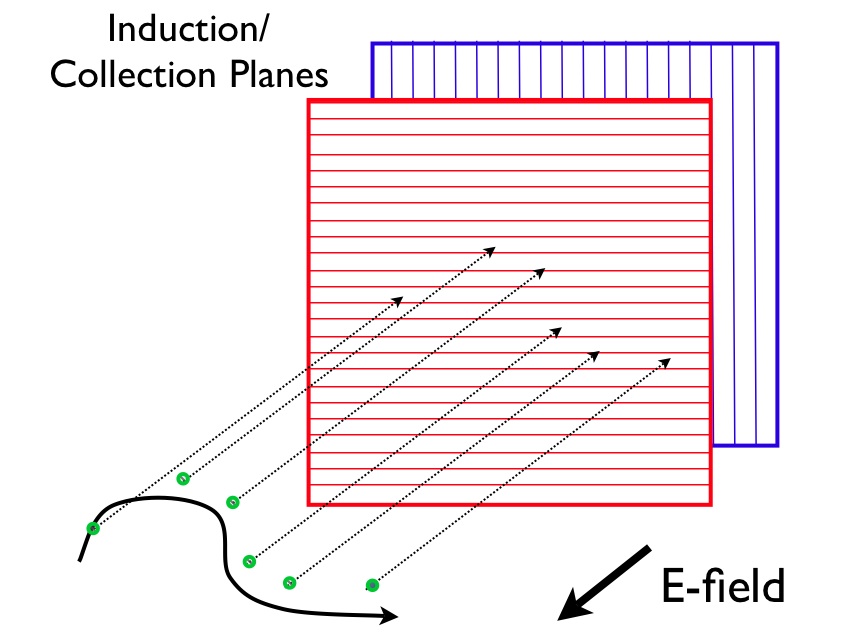}  
   \caption{Schematic diagram of TPC operation.  Ionization is drifted along an electric field to multiple planes of readout wires, each of which is instrumented with low-noise preamplifiers and fast digitization electronics.  Position of the wire within the plane, and knowledge of drift times, allow event spatial reconstruction.}
   \label{fig:tpc}
\end{figure}

The LArTPC technique has been around for several decades, with pioneering work done as part of the ICARUS experiment \cite{Rubbia, ICARUS}.  A wire chamber is placed in highly-purified liquid argon, and an electric field is created within this detector.  Neutrino interactions with the argon inside the detector volume produce ionization electrons that drift along the electric field until they reach finely segmented and instrumented anodes ($\textit{i.e.}$ - wireplanes), upon which they produce signals that are utilized for imaging and analyzing the event that occurred, as shown in Fig. $\ref{fig:tpc}$.  Applying proper bias voltages to the wireplanes, such that electrons drift undisturbed through the initial planes, allows several complementary views of the same interaction that can be combined into a three-dimensional image of the event\cite{Grids}.  Calorimetric measurements can be extracted from the pulses observed on the wireplanes.

This technique allows for very precise imaging, the resolution being dependent on several factors:  wire pitch, plane spacing, sampling rate, and electronics S/N levels.  The wire pitch is typically on the order of several millimeters, the specific value being chosen to maximize resolution without sacrificing S/N levels.  The rapid sampling rate ($\approx$5MHz) characteristic of the readout electronics, combined with the slow drift speed of ionization ($\approx$1.5mm/$\mu$s) at nominal electric field values, equates to an image resolution of fractions of a millimeter along the drift direction (which is the coordinate common to all the wireplanes of the TPC).  The technology is further made attractive in that the number of electronics channels required for the detector does not scale directly with the volume of the detector if the drift distance is increased appropriately.   This scaling feature, along with the relatively low cost of argon, makes LArTPCs an intriguing option for future massive neutrino detectors.

 %\begin{figure*}[t]
%\centering
%\includegraphics[width=135mm]{figure2.eps}
%\caption{Example of a Full Width Figure.} \label{example_figure_col2}
%\end{figure*}

While LArTPCs are an intriguing detector technology, they are not without their challenges.  One of the biggest challenges is producing and maintaining argon that is pure enough to allow the ionization to drift for the required distances.  To address this issue new filters that can cleanse the argon to the required purity levels necessary for a LArTPC experiment, and can also be regenerated when they have become saturated, have been developed\cite{Filter}.  These new filters are a necessary step along the path to massive detectors, and they have already been used by several test stands built in the U.S. with the goals of studying detector material effects on argon purity, and looking for cosmic-ray events in a LArTPC \cite{Yale}.

%%%%%%%%%%%%%%%%%%%%%%%%%%%%%%%%%%
\section{ArgoNeuT}
ArgoNeuT is a LArTPC that is currently running in the NuMI beamline at Fermilab.  The ArgoNeuT project was started in order to gain experience building and operating LArTPCs in a real beam environment, and also to collect a very interesting data sample that will be used to develop simulation and reconstruction code.  ArgoNeuT will provide a sample of neutrino interactions in liquid argon for the first time ever in the U.S., and for the first time ever in a low-energy beam.  The only previous LArTPC to operate in a neutrino beam was a 50-liter TPC built as part of the ICARUS program that ran in the WANF beam at CERN in the late 1990's \cite{50L}.  The energy of the NuMI beam (peaking at $\approx$3 GeV) is significantly lower than that of the WANF beam (mean energy of $\approx$24 GeV), making the data accumulated by ArgoNeuT particularly interesting since this is an energy-range relevant to neutrino oscillation physics.

\subsection{TPC}

\begin{figure}[htbp] %  figure placement: here, top, bottom, or page
   \centering
   \includegraphics[width=3in]{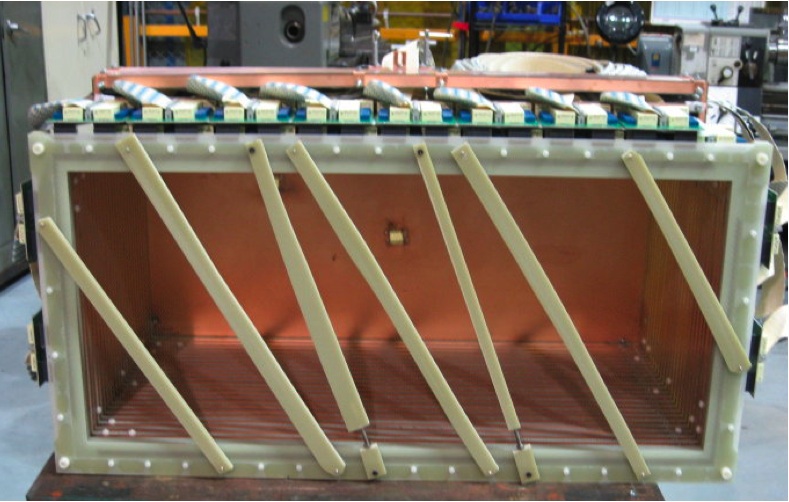}  
   \includegraphics[width=3in]{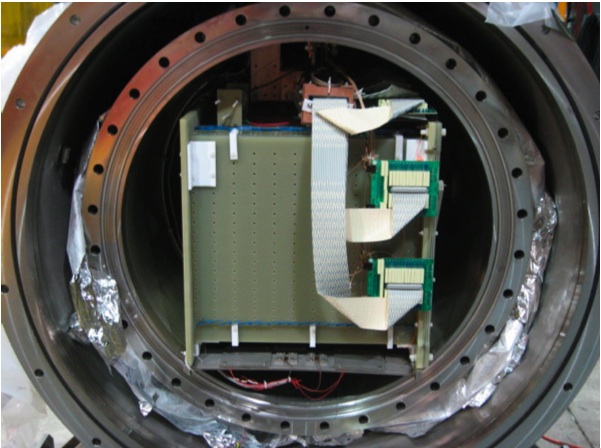}  
   \caption{The ArgoNeuT TPC.}
   \label{fig:tpc_pics}
\end{figure}

ArgoNeuT's TPC is a rectangular volume measuring 90cm x 40cm x 48cm, containing an active volume of $\approx$175 liters of liquid, that is positioned inside of a vacuum jacketed cryostat.  Figure \ref{fig:tpc_pics} depicts the TPC before and after it was inserted into the inner cryostat volume.  The TPC consists of three wireplanes, each with 4mm wire pitch.  The innermost induction plane has vertical wire orientation, but is not instrumented with readout electronics and is used primarily for pulse shaping. The middle induction plane has 240 wires oriented at +60$^{\circ}$ with respect to the horizontal beam direction, while the outermost collection plane has 240 wires oriented at -60$^{\circ}$ with respect to the horizontal beam direction.  Both the middle induction plane and the collection plane are instrumented with readout electronics.  The maximum drift distance in the TPC, from the cathode to the first induction plane is 48cm.  The operating cathode voltage of 25kV creates an electric field of 500V/cm, at which the drift speed is 1.55 mm/$\mu$s.  The inside of the TPC contains 23 field ``rings" that are 1cm in thickness, formed from machined copper-clad G10 sheets, with consecutive rings connected by four 100M$\Omega$ resistors in parallel.       

\subsection{Electronics}

\begin{figure}[htbp] %  figure placement: here, top, bottom, or page
   \centering
      \includegraphics[width=1.0in,height=1.2in]{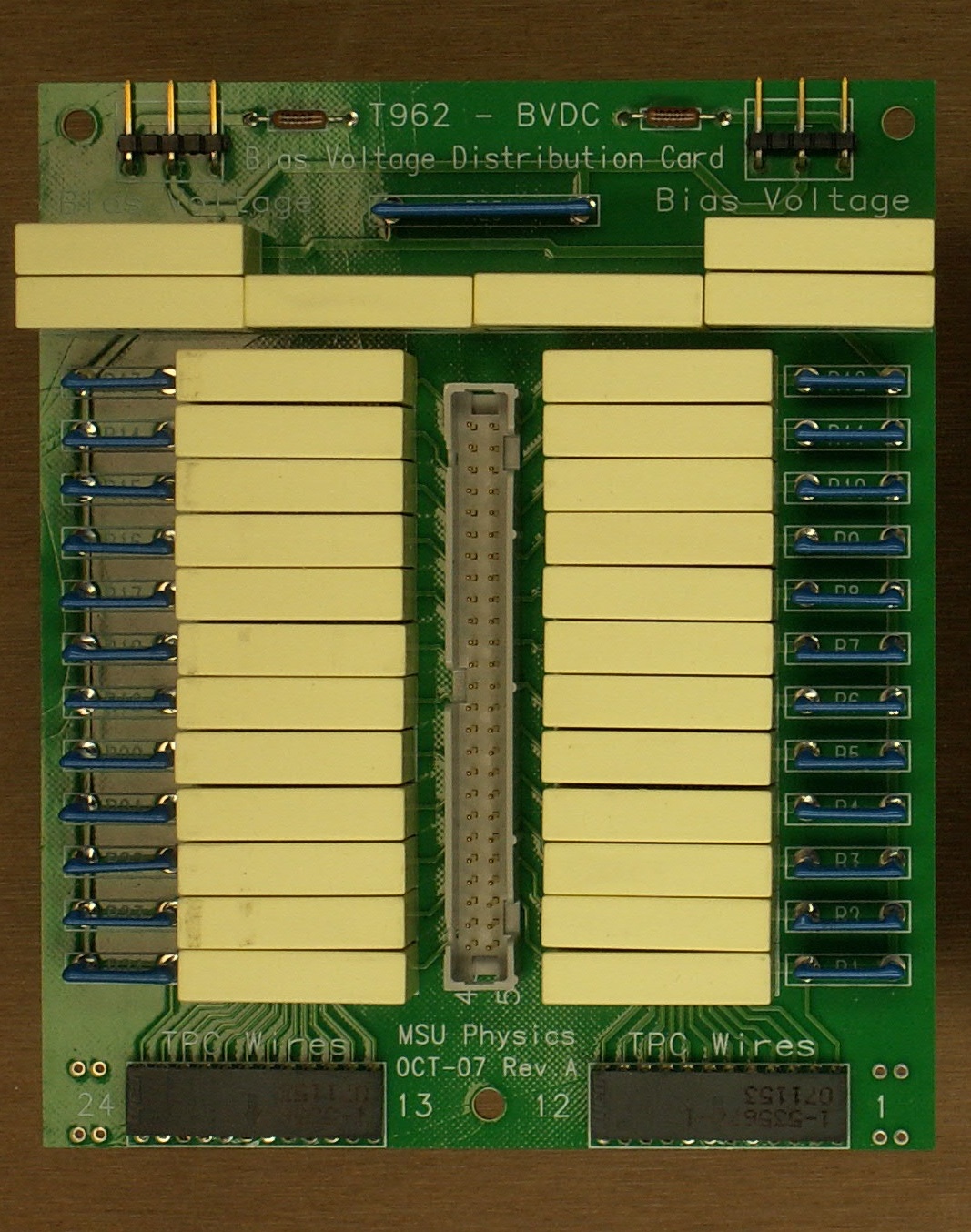}  
   \includegraphics[width=1.0in,height=1.2in]{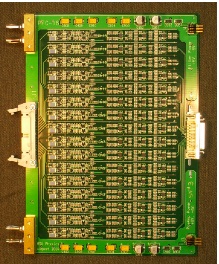}  
   \includegraphics[width=1.0in,height=1.2in]{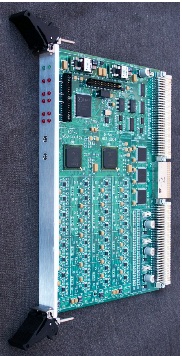}  
   \caption{Some of the components of the ArgoNeuT readout system.  Left:  Bias Voltage Distribution Card.   Center:  Preamplifier board.  Right:  ADF2 board.}
   \label{fig:electronic_pics}
\end{figure}

A custom electronic readout system has been built for the ArgoNeuT detector.  Bias voltage distribution cards (BVDCs), that provide filtered voltage to the wireplanes, are placed directly on the TPC.  Each BVDC connects to 24 TPC wires, and sends output signals to ribbon cables that connect to a custom feedthrough circuit board designed by Fermilab.  Preamplifier boards, each of which contain 16 FET preamplifiers, reside in a Faraday-cage enclosure surrounding the signal-feedthrough board flange.  The input signals to the preamplifier boards are sent through a wide bandwidth filter that removes frequencies outside of the expected range.  The amplified signals are sent to digitization boards (ADF2 cards, on loan from the D0 experiment) which sample the waveform at 5MHz (198ns/sample).  The DAQ system is triggered by a clock signal associated with the NuMI beam spill, causing each channel to begin recording 2048 ADC samples with 10-bit resolution.  The total readout time for a single trigger is $\approx$400$\mu$s, which is significantly longer than the maximum drift time of particles in the TPC (333$\mu$s) allowing for pre/post-sampling of each spill.  The pre/post-sampling information is useful for removing spurious tracks that come from outside of the beam window.  Figure \ref{fig:electronic_pics} shows several of the components of the TPC readout system.

\subsection{Cryogenics}

\begin{figure}[htbp] %  figure placement: here, top, bottom, or page
   \centering
      \includegraphics[width=2.5in]{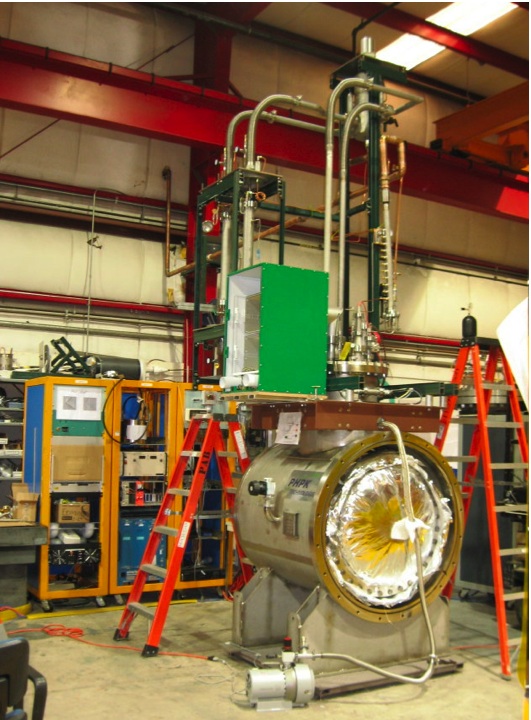}  
   \caption{ArgoNeuT's Cryogenic System.  The outer cryostat flange is removed.}
   \label{fig:cryo_pics}
\end{figure}

The main component of the ArgoNeuT cryogenic system is a 550 liter vacuum-insulated cryostat that houses the TPC and contains feedthrough ports for all of the instrumentation of the experiment.  ArgoNeuT uses a self-contained recirculation system to continually pass the liquified argon in the system through the new Fermilab style filters.  Boil-off vapor from the inner cryostat is directed vertically up to a 300W Gifford-McMahon cryocooler, where it is condensed and directed back down through one of three return paths to the inner cryostat.  Two of these return paths contain filters, while the third is a bypass line that is used during maintenance of the other pathways.  The three return lines merge before entering the cryostat, wherein they are guided down to the bottom of the liquid volume and empty through a sintered-metal cap.  Figure \ref{fig:cryo_pics} shows the ArgoNeuT detector as it looked in the summer of 2008 during a commissioning run on the surface.  In this figure the outer cryostat flange is removed, showing the inner cryostat wrapped in superinsulation.

The experiment is outfitted with numerous safety features to maintain the Oxygen Deficiency Hazard (ODH) rating of the NuMI tunnel.  All cryogenic plumbing contains relief valves that are routed to a vent line that extends up the NuMI shaft and out to the surface.  In this way any argon gas that is vented, or that might leak from a pipe, is not released into the tunnel but rather is guided out to the surface.  The outer cryostat acts as a secondary containment vessel in case the inner vessel leaks, and a ``bathtub" acts as tertiary containment in case both cryostats develop leaks.  The ``bathtub" contains ODH sensors that trigger alarms and mixing fans if a leak is detected.  A dedicated process-control system was built for ArgoNeuT that allows remote monitoring of all systems and remote control over important valves.

\subsection{Location}
  
  ArgoNeuT is currently running in the NuMI tunnel at Fermilab, where it is positioned approximately in the center of the beam, and directly upstream of the MINOS near detector.  ArgoNeuT's TPC is too small to contain the majority of muons produced in neutrino interactions from the energetic NuMI beamline.  To compensate for the information lost by particles exiting the detector, ArgoNeuT will utilize the MINOS near detector as a range stack to capture the full trajectories of these particles.  Since MINOS is magnetized there is the possibility to perform charge identification by matching a track in ArgoNeuT with a track in MINOS.

  \begin{figure}[htbp] %  figure placement: here, top, bottom, or page
   \centering
   \includegraphics[width=3in]{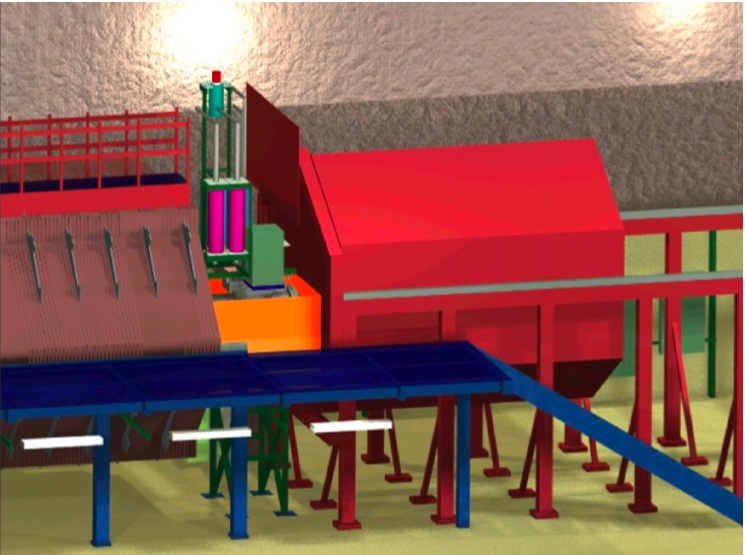}  
   \caption{Diagram of experiments in the NuMI tunnel.  ArgoNeuT is located directly upstream of the MINOS near detector, and directly downstream of the MINERvA detector.}
   \label{fig:tunnel}
\end{figure}

%%%%%%%%%%%%%%%%%%%%%%%%%%%%%%%%%%
\section{ArgoNeuT Physics}

Despite ArgoNeuT's small detector volume, it will collect a significant sample of neutrino/antineutrino interactions.  As was mentioned previously, this will be the first such sample from a LArTPC operating in the U.S., and the first sample ever in a low-energy (few GeV) neutrino beam.  There are several physics analyses that can be carried out with the ArgoNeuT data sample, with one of the most interesting coming from the several thousand charged-current quasi-elastic (CCQE) events that will be collected in antineutrino mode.  These events will be recognized by a muon created in the volume of the TPC, possibly accompanied by one or more proton tracks created by a final-state neutron.  By utilizing the MINOS near-detector to range-out muons originating from ArgoNeuT their energy can be determined, allowing a first measurement of the CCQE cross-section on argon for neutrinos in the few-GeV range.  This result will be particularly interesting for neutrino oscillation physics, since knowledge of the CCQE cross-section is crucial to oscillation analyses.  The ability of LArTPCs to see low-energy nuclear fragments created in neutrino interactions, and to determine their impact on the cross-section measurements, may also prove to be very interesting.  

Members of the ArgoNeuT collaboration are working on developing a full software environment for analyzing their data.  The software being developed will be utilized for everything from simulating neutrino interactions in a LArTPC to reconstructing the interactions starting from the raw TPC data.  Such software does not currently exist in the U.S., but is a necessary tool for any future large LArTPC detector where the statistics of the data will be greatly increased.  One of the main goals of ArgoNeuT is to utilize this new software to fully develop the $dE/dx$ particle identification technique, and to provide a measurement of the capabilities of the technology to separate electron tracks from photon tracks.  This software will be used for future LArTPC detectors, such as the MicroBooNE experiment, so early experience gained from ArgoNeuT will be important.

%%%%%%%%%%%%%%%%%%%%%%%%%%%%%%%%%%
\section{ArgoNeuT Status}

  \begin{figure}[h] %  figure placement: here, top, bottom, or page
   \centering
   \includegraphics[width=80mm]{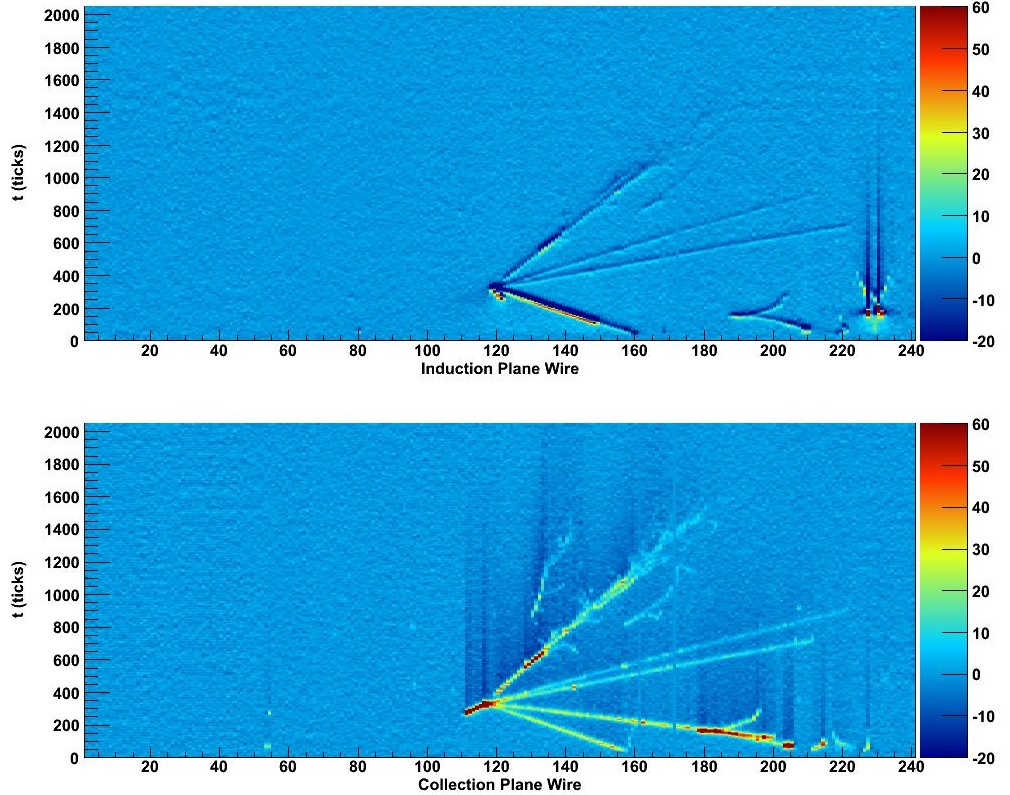}  
   \caption{Neutrino event candidate from ArgoNeuT.  The raw data for the instrumented induction and collection planes are displayed.}
   \label{fig:argoneut_event1}
\end{figure}

\begin{figure}[h] %  figure placement: here, top, bottom, or page
   \centering
   \includegraphics[width=80mm]{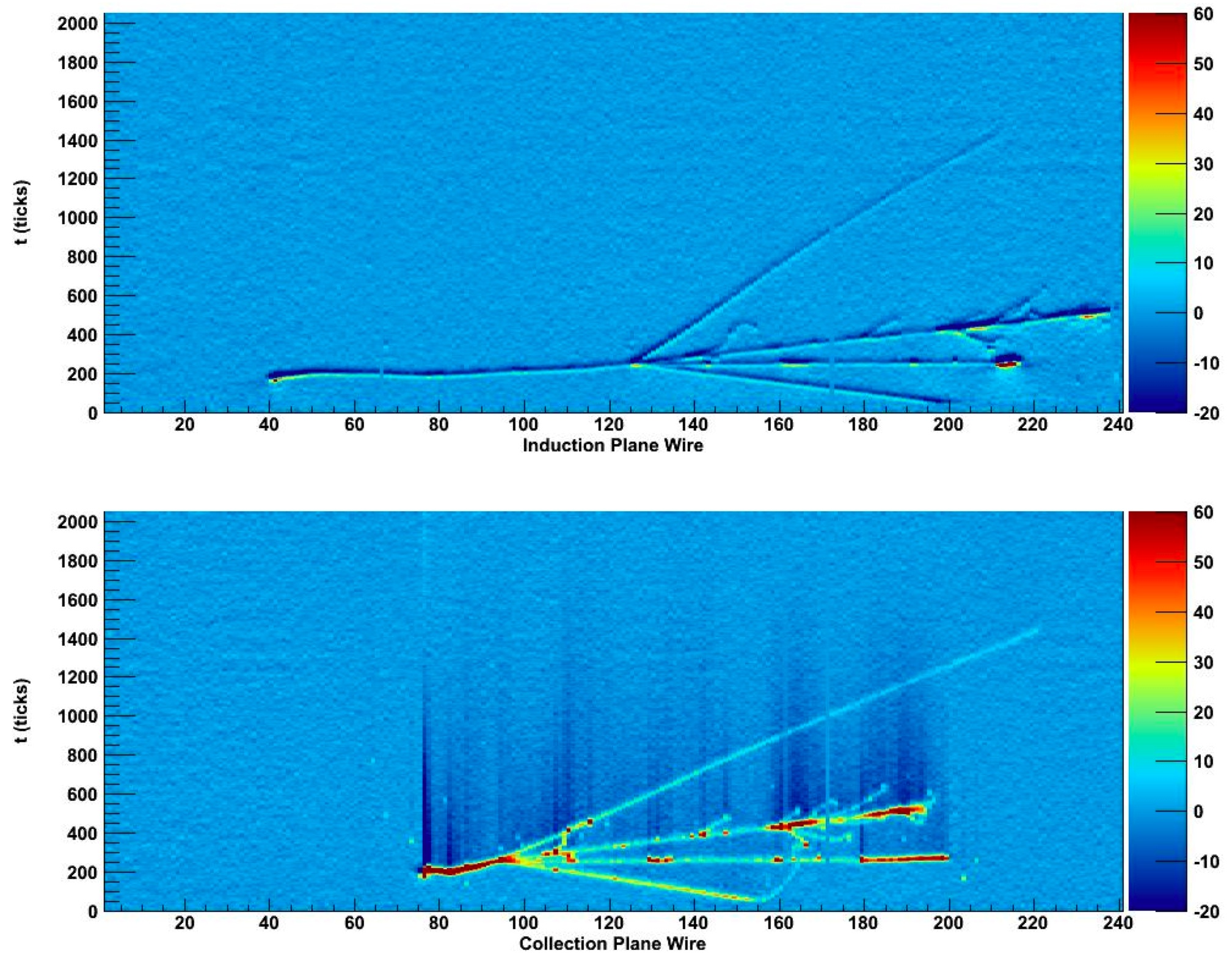}  
   \caption{Neutrino event candidate from ArgoNeuT.}
   \label{fig:argoneut_event2}
\end{figure}

\begin{figure}[h] %  figure placement: here, top, bottom, or page
   \centering
   \includegraphics[width=80mm]{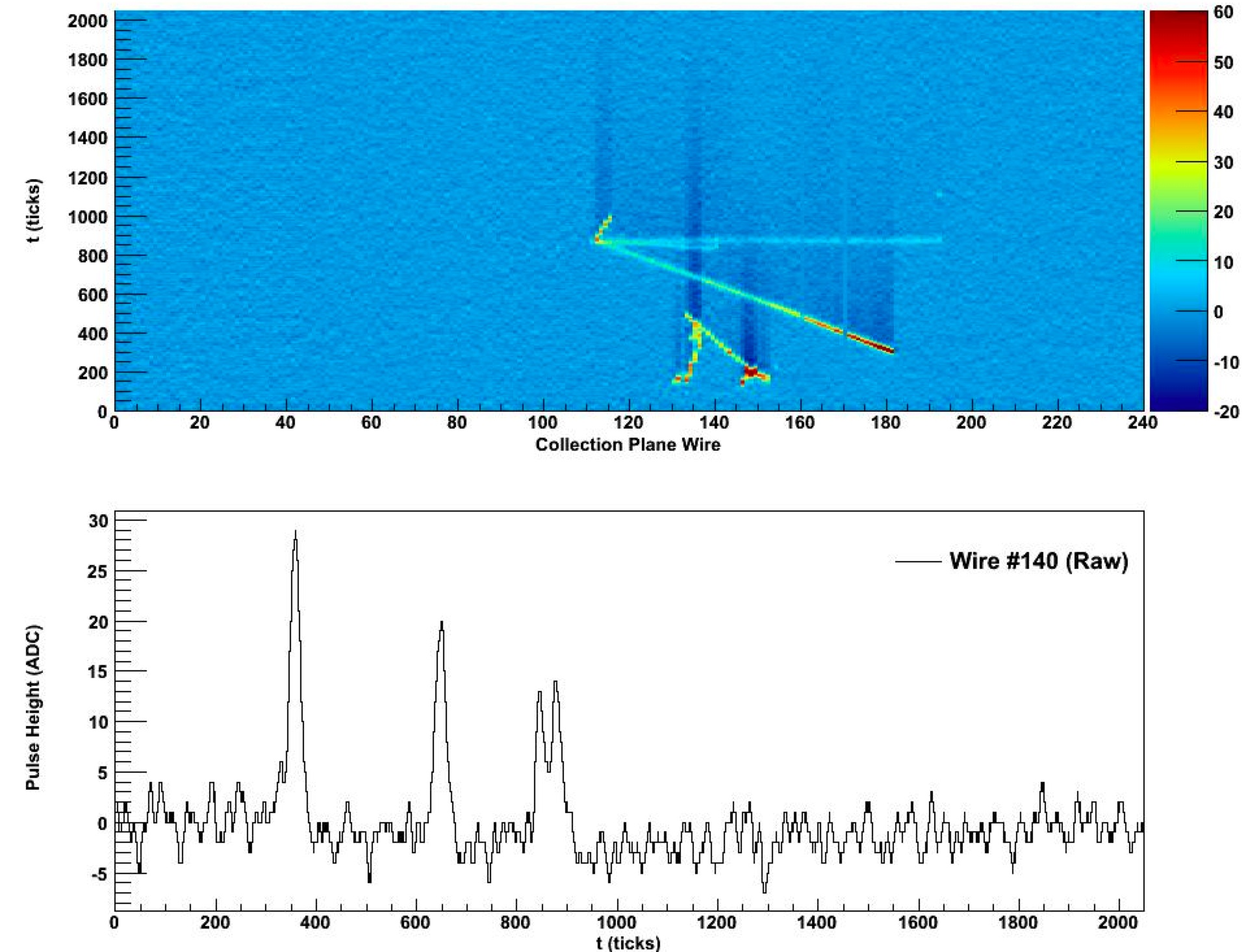}  
   \caption{Neutrino event candidate from ArgoNeuT.  The raw signal from wire $\#$140 of the collection plane is displayed.}
   \label{fig:argoneut_event3}
\end{figure}

ArgoNeuT was filled with liquid argon for the first time in its underground location in May 2009.  The initial electron lifetime was much lower than anticipated, but after several weeks of recirculating through the closed-loop filtration system the lifetime had recovered significantly.  Many neutrino events were recorded during this initial run before the summer 2009 Fermilab shutdown.  The raw data from several of these events are displayed in Figs. \ref{fig:argoneut_event1}, \ref{fig:argoneut_event2}, and \ref{fig:argoneut_event3}

Each event display depicts the information from the instrumented induction and collection plane of ArgoNeuT.   The horizontal axis is the wire number within each of the planes, while the vertical axis is the sampling time of the DAQ, which is common to both the induction and collection views.  Figure \ref{fig:argoneut_event3} depicts the collection plane view, and also shows the raw pulse information for a particular wire ($\#$140) of the collection plane.  There are three clear pulses visible in this wire, with the third pulse containing a double-peak that indicates the presence of two closely spaced tracks. 

The collaboration is currently developing algorithms for analyzing the TPC data.   The dark vertical bands visible in the collection view of Figures \ref{fig:argoneut_event2} and \ref{fig:argoneut_event3} are a result of a baseline shift in the ADCs after recording energetic hits.  Fourier deconvolution  can be performed on the raw data to accurately remove this baseline shift and return the true waveform.  Several hit finding methods are being considered to isolate the important sections of the waveforms recorded after each trigger.  Extracting overlapping hits, such as those depicted in Figure \ref{fig:argoneut_event3}, will provide information about the two-track separation achievable in these detectors.

ArgoNeuT will continue to take data through the Fall of 2009 and into early 2010, primarily in antineutrino mode.

%Figures may be one column (e.g.~Figure~\ref{example_figure}) 
%or span the paper width (e.g.~Figure~\ref{example_figure_col2}).
%Labels should be used to refer to the figures in the tex file.
%The figure caption should appear below the figure.  
%Color figures are encouraged; however note that some colors, such as 
%pale yellow, will not be visible when printed.  
%Lettering in figures should be large enough to be visible.
%Filenames of figures in the tex file should be numbered
%consecutively, e.g. figure1.ps, figure2a.ps, figure2b.ps, figure3.ps, etc.

%%%%%%%%%%%%%%%%%%%%%%%%%%%%%%%%%%

%\section{Paper Submission}

%Authors should submit their papers to the ePrint arXiv 
%server\footnote{http://arxiv.org/help} 
%after verifying that it is processed correctly by the LaTeX processor.
%Please submit the source code, the style files 
%(revsymb.sty, revtex4.cls, slac\_two.rtx) 
%and any figures; 
%these should be self-contained to generate the paper from source.  

%It is the author's responsibility to ensure that the papers are 
%generated correctly from the source code at the ePrint server. 
%After the paper is accepted by the ePrint server, please verify that
%the layout in the resulting  PDF file conforms to the guidelines 
%described in this document.
%Finally, contact the organizers of DPF-2009 and your parallel session conveners 
%(see http://www.dpf2009.wayne.edu) with the ePrint number of the paper; 
%the deadline to do this is 2~October~2009.

%%%%%%%%%%%%%%%%%%%%%%%%%%%%%%%%%%
\begin{acknowledgments}
The author would like to acknowledge the support staff at Fermilab for their invaluable contributions in the planning and construction of ArgoNeuT.  Also, the Department of Energy and the National Science Foundation.
\end{acknowledgments}

\bigskip % extra skip inserted

\end{document}